\documentclass[%
 aip,
 amsmath,amssymb,
 reprint,%
]{revtex4-1}

\usepackage{graphicx}
\usepackage{dcolumn}
\usepackage{bm}

\usepackage[utf8]{inputenc}
\usepackage[T1]{fontenc}
\usepackage{mathptmx}

\usepackage{color}
\usepackage{ulem}

\begin{document}

\preprint{AIP/123-QED}

\title{Hard x-ray photoemission study on strain effect in LaNiO$_3$ thin films}

\author{K. Yamagami}
 \email{kohei.yamagami@oist.jp (at present)}
 \affiliation{Institute for Solid State Physics, The University of Tokyo, Kashiwa, Chiba 277-8581, Japan.}
 \affiliation{Okinawa Institute of Science and Technology Graduate University, Onna-son, Okinawa 904-0495, Japan.}
\author{K. Ikeda}
 \affiliation{Institute for Solid State Physics, The University of Tokyo, Kashiwa, Chiba 277-8581, Japan.}
\author{A. Hariki}
\affiliation{Department of Physics and Electronics, Graduate School of Engineering, Osaka Prefecture University, Sakai, Osaka 599-8531, Japan.}
\author{Y. Zhang}
 \affiliation{Graduate School of Material Science, University of Hyogo, Kamigori, Hyogo 678-1297, Japan.}
\author{A. Yasui}
\author{Y. Takagi}
\affiliation{Japan Synchrotron Radiation Research Institute, Sayo, Hyogo 679-5198, Japan.}
\author{Y. Hotta}
\affiliation{Graduate School of Material Science, University of Hyogo, Kamigori, Hyogo 678-1297, Japan.}
\author{T. Katase}
\author{T. Kamiya}
 \affiliation{Laboratory for Materials and Structures, Institute of Innovative Research, Tokyo Institute of Technology, Midori, Yokohama, Kanagawa 226-8503, Japan.}
\author{H. Wadati}
\affiliation{Institute for Solid State Physics, The University of Tokyo, Kashiwa, Chiba 277-8581, Japan.}
\affiliation{Graduate School of Material Science, University of Hyogo, Kamigori, Hyogo 678-1297, Japan.}
\affiliation{Institute of Laser Engineering, Osaka University, Suita, Osaka 565-0871, Japan.}

\date{\today}

\begin{abstract}
The strain effect from a substrate is an important experimental route to control electronic and magnetic properties in transition-metal oxide (TMO) thin films.
Using hard x-ray photoemission spectroscopy, we investigate the strain dependence of the valence states in LaNiO$_{3}$ thin films, strongly correlated perovskite TMO, grown on four substrates:~LaAlO$_{3}$, (LaAlO$_{3}$)$_{0.3}$(SrAl$_{0.5}$Ta$_{0.5}$O$_{3}$)$_{0.7}$, SrTiO$_{3}$, and DyScO$_{3}$.
A Madelung potential analysis of core-level spectra suggests that the point-charge description is valid for the La ions while it breaks down for Ni and O ions due to a strong covalent bonding between the two.
A clear x-ray photon-energy dependence of the valence spectra is analyzed by the density functional theory,
which points to a presence of the La 5$p$ state near the Fermi level.
\end{abstract}

\maketitle
Transition-metal perovskite oxides (TMO) and their thin-film heterostructures host various physical properties with potential applications toward novel electronic device~\cite{TMOreview}.
Among them, LaNiO$_{3}$ (LNO) is attracting much attentions;~initially promoted by
theoretical prediction on possible superconductivity in LaNiO$_{3}$/LaAlO$_{3}$ heterostructure inheriting two-dimensional electronic structure of high-$T_{c}$ cupartes~\cite{analogyI,analogyII,analogyIII}, and more recently, by magnetic and electronic responses to strain and dimensional control by film thickness, such as metal-insulator transition (MIT)~\cite{RNOphasediagram,RNOthisupReviewI,RNOthisupReviewII,RNOthisupReviewIII,MITI,MITII,MITIII,MITIV,MITV,LNOLAOthidepARPES}, charge disproportionation (CD)~\cite{CDI,CDII,CDIII,CDIV,CDV} and spin-density waves~\cite{SDWI,SDWII,SDWIII}.

Bulk $R$NiO$_{3}$ ($R$ = Lu--Pr) show a MIT accompanied by a CD (2Ni$^{3+}\rightarrow $ Ni$^{3+\delta}$+Ni$^{3-\delta}$), in which O 2$p$ states interpose owing to a small charge-transfer energy between Ni 3$d$ and O 2$p$ states~\cite{NNORaman,NNORSXS,LuNOPressure,RENOXAFS,ENOsolgelXPS}.
LNO, unique among $R$NiO$_{3}$, is a paramagnetic metal at all temperatures.
Though the absence of MIT could be due to its large band width, an enhanced electron mass renormalization ($\sim$6$m_{0}$) suggests the importance of the electronic correlation in LaNiO$_{3}$~\cite{LNOLAOthidepARPES,LNOthisubOptial}.
The dimensional control and strain effects unveil the correlated physics in LaNiO$_{3}$.
The former amplifies the correlation effect by splitting the degenerated Ni $e_{g}$ orbitals, leading to e.g., an orbital-selective Mott transition with an orbital polarization in the Ni $e_g$ manifolds~\cite{Medici}.
The latter, on the other hand, controls the Ni band widths.
Indeed it is reported that the tensile strain sensitively increases the resistivity of LaNiO$_3$ thin films.
Although the film shows metallic conduction, the Hall coefficient decreases under tensile strain, which is closely linked with the localization of the Ni $e_g$ electrons~\cite{MITI,LNOfilmI,LNOfilmII}.
A theoretical study found that the volume of the electron pocket near the Fermi level ($E_{F}$) increases w.r.t the hole one as the strain changes from compressive to tensile~\cite{LNOfilmI}, leading a reconstruction of the density of states near $E_F$.
Thus it is of fundamental importance to understand the evolution of charge states of Ni, O, and La ions with the strain effects, that is the scope of this study.

\begin{table*}
 \caption{Epitaxial structure of LNO films on studied substrates. The in-plane lattice mismatch is evaluated by $\Delta a/a$ = ($a_{\rm{sub}}$ - $a_{\rm{LNO}}$)/$a_{\rm{LNO}}$ $\times$ 100, where $a_{\rm{sub}}$ and $a_{\rm{LNO}}$ are the pseudo-cubic lattice constants of each substrate and LNO bulk (3.8377 \AA), respectively. The lattice constants for in-plane ($a_{\rm{film}}$) and out-of-plane ($c_{\rm{film}}$) of LNO films were estimated from x-ray reciprocal space maps in Fig.S1 of the Supprelental Materials. The epitaxial strain is evaluated by $\varepsilon_{xx}$ = ($a_{\rm{film}}$ - $a_{\rm{bulk}}$)/$a_{\rm{bulk}}$ $\times$ 100 for in-plane strain and $\varepsilon_{zz}$ = ($c_{\rm{film}}$ - $c_{\rm{bulk}}$)/$c_{\rm{bulk}}$ $\times$ 100 for out-of-plane strain, respectively.}
  \vspace{5mm}
  \begin{tabular}{cccccccc}
    \hline
    \hline
    \ \ \ \ \mbox{Substrate}\ \ \ \ &\ \ \ \ \ \mbox{$a_{\rm{sub}}$ (\AA)}\ \ \ \ &\ \ \ \ \mbox{$\Delta a/a$}\ \ \ \ &\ \ \ \ \mbox{$a_{\rm film}$ (\AA)}\ \ \ \ &\ \ \ \ \mbox{$c_{\rm film}$ (\AA)}\ \ \ \ &\ \ \ \ \mbox{$c_{\rm film}/a_{\rm film}$}\ \ \ \ &\ \ \ \ \mbox{$\varepsilon_{xx}$(\%)}\ \ \ \ &\ \ \ \ \mbox{$\varepsilon_{zz}$(\%)}\ \ \ \ \\
    \hline
    \mbox{LAO} & \mbox{3.79} & \mbox{-1.50} & \mbox{3.7986} & \mbox{3.8916}& \mbox{1.025}& \mbox{-1.019}&\mbox{+1.405} \\
    \mbox{LSAT} & \mbox{3.88} & \mbox{+1.10} & \mbox{3.8645} & \mbox{3.8158} & \mbox{0.987} & \mbox{+0.386} & \mbox{-0.291} \\
    \mbox{STO} & \mbox{3.91} & \mbox{+1.75} & \mbox{3.8937} & \mbox{3.7943}& \mbox{0.975}& \mbox{+1.460}&\mbox{-1.130} \\
    \mbox{DSO} & \mbox{3.94} & \mbox{+2.66} & \mbox{3.9294} & \mbox{3.7951}&\mbox{0.966} &\mbox{+2.389} &\mbox{-1.110} \\
    \hline
    \hline
  \label{TableI}
  \end{tabular}
\end{table*}
\begin{figure*}
\begin{center}
\includegraphics[width=15cm]{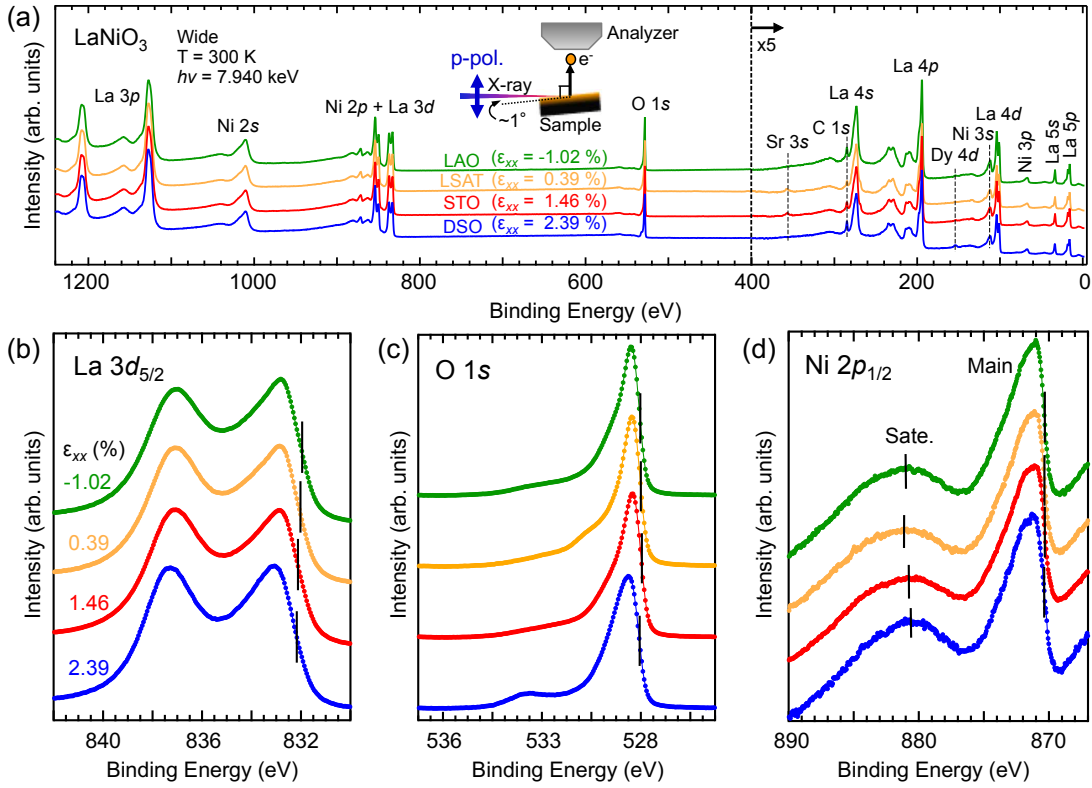}
\end{center}
\vspace{-6mm}
\begin{center}
\caption{(Color online) HAXPES spectra of LNO thin films epitaxially grown on LAO, LSAT, STO, and DSO substrates. (a) Wide-$E_B$ range, (b) La 3$d_{5/2}$, (c) O 1$s$, and (d) Ni 2$p_{1/2}$ spectra are shown for different in-plane strains $\varepsilon_{xx}$. In the wide spectra, the signals from substrates such as Sr 3$s$ and Dy 4$d$ are also shown as vertical dashed lines. In La 3$d_{5/2}$, O 1$s$, and Ni 2$p_{1/2}$ core-levels, the midpoints of the lower-$E_{B}$ slopes are depicted by vertical black solid bars. The peak positions of Ni 2$p_{1/2}$ satellite structures determined by the gaussian fitting are also shown. In the panel (a), the experimental geometry in the present HAXPES measurements is illustrated.}
\label{Fig.1}
\end{center}
\end{figure*}
Soft x-ray photoemission spectroscopy (SXPES) has been employed to study LNO thin films~\cite{LNOSTOthiSXPESXAS,LNOSTOthiSXARPES,LNOSTOthiSWPES,LNOLAOthidepSXPES,LNONdSTOthidepSXAPPES,LNOsubdepARPES,LNOLAOthidepSXAPRES}.
The SXPES spectra, for example, present a momentum dependence of the mass renormalization in three-dimensional band structures~\cite{LNOSTOthiSXARPES}.
However SXPES is surface sensitive since the photoelectron inelastic mean free path (IMFP) in SXPES is 5-23 \AA~\cite{IMFPI,IMFPII,IMFPIII}.
Thus, SXPES suffers from obstacles, e.g.,~surface roughness due to by-hand cleaving of the sample or air pollution, leading to a surface reconstruction and carrier doping by impurities.
Besides, the small detection depth poses a difficulty to study buried layers and interfaces of the epitaxial heterostructure.
In contrast, hard x-ray photoemission spectroscopy (HAXPES) measures the bulk electronic states since IMFPs are 30-150 \AA\ for photon energies ($h\nu$) above 2~keV.
The large IMFPs allows us to avoid the surface effects and probe the signals from interfaces.
The MIT transition in LNO thin films by thickness control was investigated in HAXPES~\cite{LNOthisubdepHAXPES}.
Since the lack of apical oxygen to the NiO$_{6}$ octahedron on the surface of LNO thin films produces a strain dependence of the energy splitting of Ni $e_{g}$ orbitals, being different from that of its bulk~\cite{LNOfilmII}, HAXPES is an effective tool to investigate the strain dependence of the bulk electronic structure of LNO films.

In this paper, we study the strain dependence of the electronic structure in LNO thin films using valence and core-level HAXPES.
We compare thin films grown on four substrates:~LaAlO$_{3}$ (LAO), (LaAlO$_{3}$)$_{0.3}$(SrAl$_{0.5}$Ta$_{0.5}$O$_{3}$)$_{0.7}$ (LSAT), SrTiO$_{3}$(STO), and DyScO$_{3}$ (DSO).
By a systematic analysis of the core-level shift using a Modeling-potential calculation and density functional theory (DFT) calculation, the valence state of La, Ni and O ions are investigated.
The valence HAXPES spectra near $E_{F}$ show a sharp evolution with the tensile strain.
The contribution of the semicore La 5$p$ to the valence spectra due to its large cross-section in hard x-ray regime is identified by a DFT analysis taking the photoionization cross-section into account.

LNO thin films with a thickness of $\sim$25 nm were fabricated by pulsed laser deposition on the substrates of (001)-oriented LAO with in-plane lattice mismatch ($\Delta a/a$) of -1.50\%, LSAT with $\Delta a/a$ = +1.10\%, STO with $\Delta a/a$ = +1.75\%, and DSO with $\Delta a/a$ = +2.66\% at a substrate temperature of 700 $^{\circ}$C under high O$_2$ pressure of 25 Pa, and the crystallinity was confirmed by the low energy electron diffraction.
Note that the lattice mismatches of GdScO$_3$ ($\Delta a/a$ = +3.2\%) and NdScO$_3$ ($\Delta a/a$ = +4.2\%) are too large, the lattice relaxation occurs for LNO films on these substrates.
Therefore, the largest tensile strain can be applied for LNO films on DyScO$_3$ ($\Delta a/a$ = +2.4\%).
Table~\ref{TableI} shows the structural information of LNO thin films.
The anisotoropy ($c_{\rm film}/a_{\rm film}$) between in-plane ($a_{\rm film}$) and out-of-plane ($c_{\rm film}$) lattice constants due to the tensile-strain is determined by the X-ray diffraction [see Fig.S1 in supplemental materials].
The in-plane strain ($\varepsilon_{xx}$ [\%]) from the substrate is defined as $\varepsilon_{xx}=(a_{\rm film}-a_{\rm bulk})/a_{\rm bulk}\times100$, where $a_{\rm bulk}$ denotes the lattice constants of LNO in bulk (3.8377 \AA).
The positive (negative) $\varepsilon_{xx}$ corresponds to tensile (compressive) strain.
We have confirmed the metallic conduction of LNO films under compressive to tensile strain.
HAXPES measurements were performed using a Scienta R4000 analyzer of SPring-8 BL47XU~\cite{BL47XU}.
As show in the inset of Fig.~\ref{Fig.1}(a), the hard x-ray of $h\nu$ = 7.940 keV with p-polarization (p-pol.) are irradiated to the sample with the gracing angle $\sim$1$^{\circ}$, and the analyzer is set to nearly perpendicular to the film surface.
We emphasis that since the thickness is about five times the effective attenuation length of 55 \AA, the substrate intensity is reduced to $e^{-5}\sim0.007$, dominantly probing the signal from LNO thin films~\cite{LNOthisubdepHAXPES}.
On the other hand, SXPES measurements were carried out using a monochromatized Al $K\alpha$ source ($h\nu$ = 1.487 keV) by a PHI5000 VersaProbe system (ULVAC-PHI Inc.).
The sample was fixed using a conducting tape and a silver paste to ensure the electrical conductivity.
The binding energy ($E_{B}$) was calibrated using deposited Au, and the energy resolution was determined at $\sim$250 meV for 7.940 keV and $\sim$500 meV for 1.487 keV respectively.
All measurements were performed at 300 K.

\begin{figure}
\includegraphics[width=8.6cm]{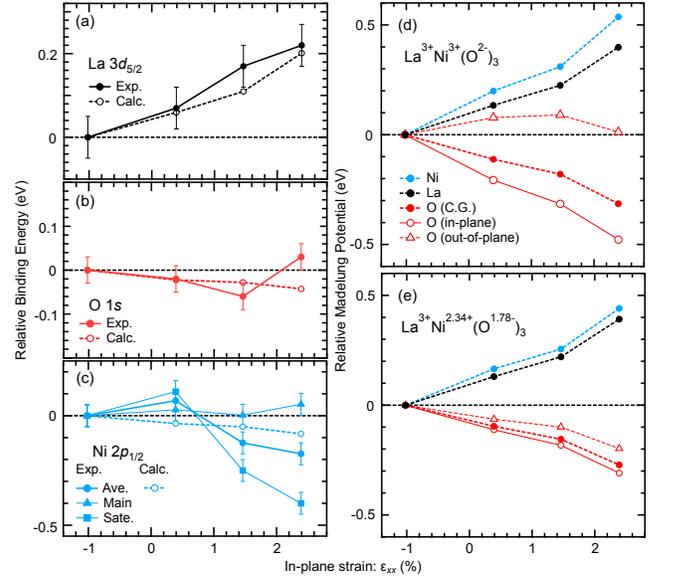}
\vspace{-6mm}
\begin{center}
\caption{(Color online) (a-c) The relative binding energy shift ($\Delta E_{B}$) of La 3$d_{5/2}$, O 1$s$, and Ni 2$p_{1/2}$ core-levels in LNO thin films as a function of $\varepsilon_{xx}$. In Ni 2$p_{1/2}$, the energy shifts of main, satellite and their averaged value are plotted.
The energy shifts obtained by the LDA calculation are shown together.
(d-e) The Madelung potential shifts ($\Delta V_{M}$) calculated for the point-charge models of La$^{3+}$Ni$^{3+}$O$_{3}^{2-}$ and  La$^{3+}$Ni$^{2.34+}$O$_{3}^{1.78-}$.}
\label{Fig.2}
\end{center}
\end{figure}
\begin{figure}
\includegraphics[width=8.6cm]{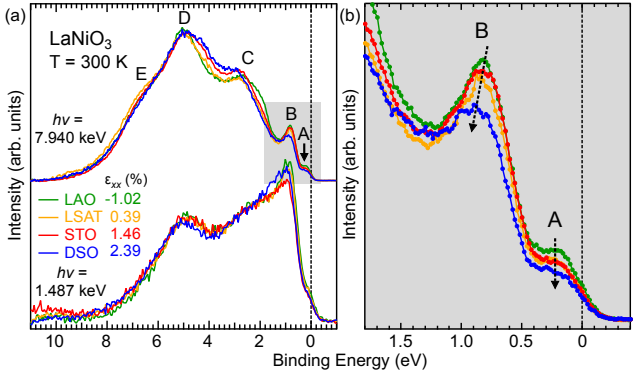}
\vspace{-6mm}
\begin{center}
\caption{(Color online) (a) Valence-band spectra at $h\nu$ = 7.940 keV and 1.487 keV for LNO thin film on different substrates. All spectra are normalized by the area of valence-band after subtracted by shierly-type background. The main spectral features are labeled as A-E. (b) The low-$E_B$ valence spectra corresponding to the gray-shadow region in the panel~(a). The dashed arrows in $h\nu$ = 7.940 keV show $\varepsilon_{xx}$-dependence of the peaks A and B.}
\label{Fig.3}   
\end{center}
\end{figure}

We first investigate the $\varepsilon_{xx}$-dependence of the core-level spectra.
The spectra in wide energy range indicate a good quality of our LNO thin films;~a negligibly small signal of C 1$s$ peak ($E_{B}$ = 285 eV) is due to a possible contaminant on the surface, and the signals of Sr 3$s$ ($E_{B}$ = 359 eV) and Dy 4$d$ ($E_{B}$ = 154 eV) are from the substrates [Fig.~\ref{Fig.1}(a)].
To estimate the $\varepsilon_{xx}$-dependence of the core-level shift, we analyze La 3$d_{5/2}$ and O 1$s$ core levels, shown in Figs.~\ref{Fig.1}(b) and (c).
Since the Ni 2$p_{3/2}$ is overlapped with La 3$d_{3/2}$, Ni 2$p_{1/2}$ core level is analyzed for Ni ions, see Fig.~\ref{Fig.1}(d).
The midpoint of the low-$E_{B}$ slopes in La 3$d_{5/2}$ core-level shifts to higher $E_{B}$ side about $\sim$200 meV from $\varepsilon_{xx}=$ -1.02 \% (on LAO) to $\varepsilon_{xx}=$ 2.39 \% (on DSO), while those of O 1$s$ and Ni 2$p_{1/2}$ core-levels shift to the lower $E_{B}$ side.
These core-level shifts ($\Delta E_{B}$) are summarized in Figs.~\ref{Fig.2}(a)-(c).
The $\Delta E_{B}$ of Ni 2$p_{1/2}$ is estimated by averaging the shifts of the main line and the satellite structure, see the Supplemental Materials for details.

To support the $\Delta E_B$ values estimated from the HAXPES spectra, we performed a DFT simulation for (bulk) LNO with lattice constants in Table~\ref{TableI} using WIEN2K package~\cite{WIEN2K}.
The $\Delta E_B$ in the DFT results in Figs.~\ref{Fig.2}(a)-(c) shows a reasonable agreement with the experimental estimate.
The $\Delta E_{B}$ w.r.t the chemical potential $\mu$ in core-level photoemission spectra is given by~\cite{Chemicalshift},
\begin{align}
\Delta E_{B} = \Delta\mu-K\Delta Q+\Delta V_{M}-\Delta E_{R}. \notag
\end{align}
Here, $\Delta\mu$ is the change in the chemical potential, $\Delta Q$ is the change in the number of valence electrons on the atom, and $K\Delta Q$ gives the so-called chemical shift ($K$ is a constant).
$\Delta V_{M}$ is the change in the Madelung potential, and $\Delta E_{R}$ is the change in the screening of the core hole due to conduction electrons.
With the in-plane strains $\varepsilon_{xx}$, both $\mu$ and $V_{M}$ can be varied since the band filling could be changed.
This was demonstrated for La$_{0.6}$Sr$_{0.4}$MnO$_{3}$ thin film grown on different substrates where $\Delta V_{M}$ is $\sim$200 meV according to the point-charge model analysis~\cite{Madelung}.
\begin{table*}
 \caption{The values of $\beta$ and $d\sigma/d\Omega$ at 7.940 keV and 1.487 keV~\cite{Cross-Section1,Cross-Section2,Cross-Section3}. Because of the parallel geometry with the analyzer along the electric field vector [see the inset of Fig.~\ref{Fig.1}(a)], we set to $\theta$ = 0$^{\circ}$ ($\varphi$ undefined), which means the absent of the third term in equation of $d\sigma/d\Omega$. Where, $\theta$ is the angle between the electrical field and the momentum of the photoelectron and $\varphi$ is the angle between the photon momentum vector and the projection of the photoelectron momentum vector on the plane perpendicular to the electrical field vector and containing the photon momentum vector.}
  \vspace{5mm}
  \begin{tabular}{ccccc}
    \hline
    \hline
    \ \ \ \ \ \ \ \ \ \ \ \ \ \ \ \ \mbox{$h\nu$}\ \ \ \ \ \ \ \ \ \ \ \ \ \ \ \ &\ \ \ \ \ \ \ \ \ \ \ \ \ \ \ \  \ \ \ \ \ \ \ \ \ \ \ \ \ \ \ \ &\ \ \ \ \ \ \ \ \ \ \ \ \ \ \ \ \mbox{7.940 keV}\ \ \ \ \ \ \ \ \ \ \ \ \ \ \ \ &\ \ \ \ \ \ \ \ \ \ \ \ \ \ \ \ \mbox{7.940 keV} \ \ \ \ \ \ \ \ \ \ \ \ \ \ \ \ &\ \ \ \ \ \ \ \ \ \ \ \ \ \ \ \ \mbox{1.487 keV}\ \ \ \ \ \ \ \ \ \ \ \ \ \ \ \ \\
    \hline
    \mbox{Atomic} & & \mbox{$d\sigma/d\Omega$ (p-pol.)} &  \mbox{$d\sigma/d\Omega$ (s-pol.)} & \mbox{$d\sigma/d\Omega$}\\
    \mbox{subsell} & \mbox{$\beta$} & \mbox{($10^{-4}$ kb)} & \mbox{($10^{-4}$ kb)} & \mbox{($10^{-2}$ kb)} \\
    \hline
    \mbox{Ni 3$d$}&\mbox{0.37}&\mbox{1.21}&\mbox{1.07}&\mbox{5.20}\\
    \mbox{O 2$p$}&\mbox{0.10}&\mbox{0.112}&\mbox{0.128}&\mbox{0.47}\\
    \mbox{La 5$p$}&\mbox{1.48}&\mbox{93.1}&\mbox{11.1}&\mbox{17.3}\\
    \mbox{La 5$d$}&\mbox{0.91}&\mbox{8.98}&\mbox{3.46}&\mbox{7.06}\\
    \hline
    \hline
  \label{TableII}
  \end{tabular}
\end{table*}
\begin{figure}
\includegraphics[width=8cm]{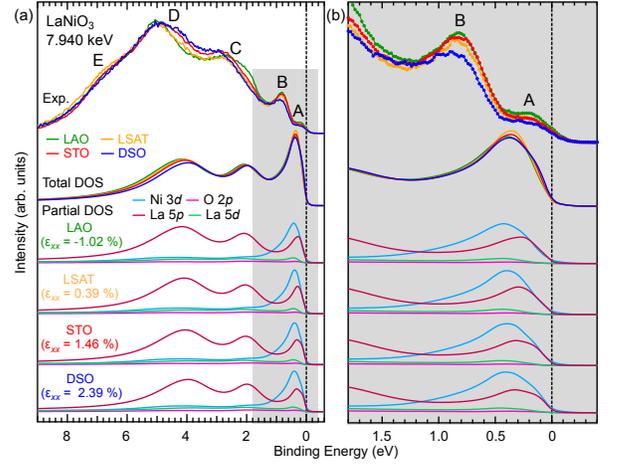}
\vspace{-6mm}
\begin{center}
\caption{(Color online) Upper two spectra are experimental and calculated total density of states (DOS) by the LDA calculations valence spectra at $h\nu$ = 7.940 keV with p-pol. configuration. The calculated spectra were obtained by multiplying the Ni 3$d$, O 2$p$, La 5$p$, and La 5$d$ partial DOS which already taking their respective $d\sigma/d\Omega$ into account, as shown in Table~\ref{TableII}. The bottom spectra shows the detail of the PDOS weights of each elements. All calculated spectra are multiplied by a Fermi Dirac function at 300 K after the convolutions with an energy resolution of 250 meV expressed in Gaussian and an energy-dependent Lorentzian broadening of 200|$E_{B}$| meV (Ref.~[\onlinecite{BroadenI}] and [\onlinecite{BroadenII}]).}
\label{Fig.4}
\end{center}
\end{figure}

To assess $\Delta V_{M}$ for the LNO thin films, the Madelung analysis is performed by Ewald's method~\cite{Ewalds} following Ref.[\onlinecite{Madelung}].
We simulated two models with different valence states for constituting ions:~La$^{3+}$Ni$^{3+}$O$_{3}^{2-}$ and La$^{3+}$Ni$^{2.34+}$O$_{3}^{1.78-}$.
The later model is suggested by the configuration-interaction cluster model simulation, taking the effects of covalency between the Ni and O atoms into account~\cite{LNOcluster}.
Figures~\ref{Fig.2}(d) and (e) show the relative $\Delta V_{M}$ in each ion as a function of $\varepsilon_{xx}$ in the two valence models, respectively.
We find that the $\Delta V_{M}$ values for the in-plane and out-of-plane O ions substantially differ from each other in both models though such a splitting is not observed in the O 1$s$ core-levels spectra.
Thus, for a compassion, we present the center of gravity of $\Delta V_{M}$ value for O ions as well.
In both models, $\Delta V_{M}$ values for Ni and O ions depart from the experimental data qualitatively:~a large negative slope of $\Delta V_{M}$ with the strain $\varepsilon_{xx}$ for the O ion is inconsistent with the negligibly small $\Delta E_{B}$ in the experiment.
The sign of $\Delta V_{M}$ for Ni ion is opposite to that of the experimental $\Delta E_{B}$.
Therefore, the Madelung potential analysis suggests that the covalency effect between Ni and O ions is substantial and cannot be described by  a static charge exchange between Ni and O in a simple ionic picture.
On the other hand, though its absolute value is overestimated, the Madelung potential analysis reproduces the $\varepsilon_{xx}$-dependence of $\Delta E_{B}$ for the La ion well.
Thus the point-charge picture is reasonable for La ions.

Next we investigate the tensile-strain effects on the valence-band photoemission spectra.
As shown in Fig.~\ref{Fig.3}(a), we observe five characteristic features labeled by A-E at $h\nu$ = 7.940 keV.
According to previous SXPES studies~\cite{LNOSTOthiSXPESXAS,LNOLAOthidepSXPES,LNONdSTOthidepSXAPPES}, features A and B are assigned as the anti-bonding states of Ni $e_{g}$ and Ni $t_{2g}$ orbitals hybridized with O 2$p$ orbitals, respectively.
The bonding and non-bonding states are located at around 2-8 eV corresponding to features C--E.
The valence states around $E_{B}$ = 0--4 eV, corresponding to A--C, show a systematic tensile-strain $\varepsilon_{xx}$ dependence, see Figs.~\ref{Fig.3}(a) and (b), which we will discuss later.
The spectral intensities change between $h\nu$ = 7.940 keV and 1.487 keV because of the $h\nu$ dependence of the photoionization cross-section ($d\sigma/d\Omega$)~\cite{Cross-Section1,Cross-Section2,Cross-Section3}.
Especially, the features D and E are enhanced to other features with increasing $h\nu$ from 1.487 to 7.940 keV.
To understand the $h\nu$ dependence, we evaluate the cross-section~\cite{Cross-Section1,Cross-Section2,Cross-Section3}
\begin{equation}
\frac{d\sigma}{d\Omega}=\frac{\sigma}{4\pi}[1+\beta P_{2}(\cos\theta)+(\gamma\cos^{2}\theta+\delta)\sin\theta\cos\varphi)]\notag \label{EQ1}
\end{equation}
for the present experimental geometry. 
The first two terms and the third one describes dipolar and nondipolar contributions, respectively.
Table~\ref{TableII} summarizes the dipole parameter of the angular distribution ($\beta$) and $d\sigma/d\Omega$ values for each element at 7.940 keV in and 1.487 keV.
The $d\sigma/d\Omega$ value for Ni 3$d$ and O 2$p$ orbitals decreases from 1.487 keV to 7.940 keV.
Remarkably, the ratio of $d\sigma/d\Omega$ of the La 5$p$ to Ni 3$d$ orbital is estimated to be $\sim$77 for 7.940 keV, which is substantially larger than $\sim$3.3 for 1.487 keV.
Thus the La 5$p$ contribution, if exists, can be substantially amplified to other orbitals in HAXPES. 
Recently, Takegami $et$ $al$.~\cite{LaCoO3HAXPES}~reported that the semicore La 5$p$ orbitals largely contribute to the HAXPES spectra including near-$E_F$ features in perovskite LaCoO$_3$.
Below we discuss La 5$p$ contributions in HAXPES of the LNO films.

Figure~\ref{Fig.4} shows the experimental HAXPES and DFT results for the LNO films on the studied substrates.
In the DFT results, the photoionization cross-section is included in a poor-man's way;~$d\sigma/d\Omega$ in Table~\ref{TableII} is applied to the density of states projected on local orbitals obtained within the local density approximation (LDA).
The bare LDA density of states can be found in the Supplemental Materials.
The simulated spectra reproduce the spectral features C--E reasonably well, including the $\varepsilon_{xx}$-dependence of feature D in the HAXPES spectra.
The orbital-resolved spectra indicate the features C and D in HAXPES are dominated by the La 5$p$ states.
However, the calculated spectra do not reproduce the low-energy features A and B. 
This discrepancy is not a surprising result.
As seen in the orbital-resolved spectra, the weights of the Ni 3$d$ contributions are comparable to those of La 5$p$ ones around A and B features.
It is known that an accurate description of the electronic correlation beyond LDA is required to reproduce spectral reconstruction with the in-plane strain $\varepsilon_{xx}$ on Ni 3$d$ bands in LNO films, see e.g.~Ref.~[\onlinecite{LNOSTOthiSXPESXAS}]. 
The features A and B show a clear $\varepsilon_{xx}$-dependence.
We speculate that it relates to amount of possible oxygen vacancies that, e.g., depend on annealing temperatures in preparing the thin films.
As it is demonstrated for the LaNiO$_{3-\delta}$ thin film~\cite{LNOSTOthiSXPESXAS}, the oxygen vacancies yield Ni$^{2+}$ ions and thus directly affect Ni 3$d$ spectral weights in low energies.
In addition to the suddenly increase of $\Delta E_{B}$ of O 1$s$ for $\varepsilon_{xx}=2.39\%$ [Fig.~\ref{Fig.2}(b)], the Ni $L_{2,3}$-edge x-ray absorption spectra of the studied samples, see Fig.S2 in supplemental materials~\cite{LNOLMOheteroXAS}, indicate a presence of Ni$^{2+}$ ions in the studied films and its amount increases with $\varepsilon_{xx}$.
The oxygen-vacancy effect on La 5$p$ states is, however, rather nontrivial, but needs to be taken into for the HAXPES spectra because of the large La 5$p$ cross-section [see Fig.~\ref{Fig.4}(b)].
To address the $\varepsilon_{xx}$-dependence in the HAXPES spectra, the polarization dependence~\cite{LDHAXPESI,LDHAXPESII,LDHAXPESIII,LDHAXPESIV,LDHAXPESV} would be valued since, as shown in Table~\ref{TableII}, the La 5$p$ cross-section can be largely suppressed employing the so-called s-polarization configuration ($\theta$ = 90$^{\circ}$ and $\varphi$ = 90$^{\circ}$ in the formula for $d\sigma/d\Omega$~\cite{Cross-Section1,Cross-Section2,Cross-Section3} above).
Thus, a liner-dichroism measurement in HAXPES would bring a further insight on the $\varepsilon_{xx}$-dependence of the electronic structure in the LaNiO$_3$ films, that will be an interesting future study.

In summary, we performed hard x-ray photoemission spectroscopy for LaNiO$_{3}$ thin films grown on four substrates to investigate the tensile-strain dependence of the electronic valence states.
The Madelung potential of core-level spectra reveal that La ion is mostly described as a point charge, while Ni and O ions are described by simply reducing the effective charges owing to the presence of the hybridization.
The density functional theory taking tensile-strain into account for the valence-band spectra indicates that the signal of La 5$p$ components are enhanced near $E_{F}$ due to angular distribution photoionization cross-section at hard x-ray regions.

We thanks K. Amemiya for supporting soft x-ray absorption spectroscopy measurements.
The hard x-ray photoemission spectroscopy measurements were performed at the BL47XU of SPring-8 with the approval of the Japan Synchrotron Radiation Research Institute (Proposal No. 2018A1073).
The soft x-ray absorption spectroscopy measurements at BL-16A of Photon Factory were carried out under the approval of the Photon Factory Program Advisory Committee (Proposal No. 2017G597, 2019G556).
This work was partially supported by JSPS KAKENHI (Grants No. 19H02425, 19H05824)
T. Katase was supported by PRESTO, Japan Science and Technology Agency (Grant No. JPMJPR16R1) and Grant-in-Aid for Challenging Research (Exploratory) (Grant No. 20K21075).
This work was supported (in part) by the Collaborative Research Project of Laboratory for Materials and Structures, Institute of Innovative Research, Tokyo Institute of Technology.

The supplementary materials summarize the details of the X-ray diffraction, the determination of the midpoint of the binding energy slopes, the bare LDA density of states, and the strain dependence of the soft x-ray absorption spectroscopy and epitaxial structure of LNO films grown on four substrates.

\section*{Data Availability}
The data that support the findings of this study are available from the corresponding author upon reasonable request.

\end{document}